\documentclass[a4paper,reprint]{revtex4-1}
\usepackage[T1]{fontenc}
\usepackage[utf8]{inputenc}

\usepackage{graphicx}
\usepackage[colorlinks,citecolor=blue,linkcolor=blue,urlcolor=blue]{hyperref}

\usepackage{bm}
\usepackage{siunitx}
\usepackage{chemmacros}

\begin{document}

\title{Exfoliation of Graphene and Fluorographene in Molecular and
  Ionic Liquids}

\author{Émilie Bordes}
\affiliation{Institute of Chemistry of Clermont-Ferrand, Universté
  Clermont Auvergne \& CNRS, 63000 Clermont-Ferrand, France}

\author{Joanna Szala-Bilnik}
\affiliation{Institute of Chemistry of Clermont-Ferrand, Universté
  Clermont Auvergne \& CNRS, 63000 Clermont-Ferrand, France}

\author{Agílio A. H. Pádua}
\email{agilio.padua@uca.fr}
\affiliation{Institute of Chemistry of Clermont-Ferrand, Universté
  Clermont Auvergne \& CNRS, 63000 Clermont-Ferrand, France}

\date{\today}

\begin{abstract}
  We use molecular dynamics simulation to study the exfoliation of
  graphene and fluorographene in molecular and ionic liquids, by
  performing computer experiments in which one layer of the 2D
  nanomaterial is peeled from a stack, in vacuum and in the presence
  of solvent. The liquid media and the nanomaterials are represented
  by fully flexible, atomistic force fields. From these simulations we
  calculate the potential of mean force, or reversible work, required
  to exfoliate the materials. Calculations in water and organic
  liquids showed that small amides (NMP, DMF) are among the best
  solvents for exfoliation, in agreement with experiment. We tested
  ionic liquids with different cation and anion structures, allowing
  us to learn about their solvent quality for exfoliation of the
  nanomaterials. First, a long alkyl side chain on the cation is
  favourable for exfoliation of both graphene and fluorographene. The
  presence of aromatic groups on the cation is also favourable for
  graphene. No beneficial effect was found between fluorine-containing
  anions and fluorographene. We also analysed the ordering of ions in
  the interfacial layers with the materials. Near graphene, nonpolar
  groups are found but also charged groups, whereas near
  fluorographene almost exclusively non-charged groups are found, with
  ionic moieties segregated to second layer. Therefore, fluorographene
  appears as a more hydrophobic surface, as expected.
\end{abstract}

\maketitle

\section{Introduction}

Two-dimensional (2D) nanomaterials are at the forefront of fundamental
and applied research today \cite{Ferrari:2015co}. Among the remarkable
features of 2D materials are their electronic structures: graphene
\cite{Novoselov:2004it} is a conductor; h-BN \cite{Novoselov:2005cg}
and fluorographene \cite{Nair:2010eh} are insulators; \ch{MoS2}, other
related transition-metal dichalcogenides (TMDC) and phosphorene are
semiconductors \cite{Wang:2012fa}. Assembling stacks of different 2D
materials \cite{Gao:2012et,Geim:2013hf,Novoselov:2016ik} according to
their function (as conductors, insulators or semiconductors) allows
fabrication of transistors, capacitors, sensors or optoelectronic
devices down to the thickness of atoms \cite{Lembke:2015fy}. By
relying on delicate noncovalent forces to provide contacts and
cohesion, the intrinsic properties of the component materials are
largely preserved (although there are junction effects). The ways in
which the layers assemble to form 2D heterostructures depends
therefore on noncovalent forces between unlike 2D materials, which are
not well described or understood at a fundamental level.

Liquid-phase exfoliation is one of the most promissing routes for the
production of 2D materials in large scale \cite{Nicolosi:2013ex}. Inks
of suspended 2D materials can be used to print electronic devices
\cite{Kelly:2017kl} on a variety of substracts, including flexible and
textile supports.  The interplay between interlayer and solvation
forces in liquid-based preparation routes (inks) is delicate and also
not well described at present. The interactions and interfacial layers
of molecular and ionic liquids with nanomaterials are important for
other fields of application, such as electrolytes for supercapacitors
\cite{Simon:2008dc} or for ionic-liquid gated transistors
\cite{Fujimoto:2013iq}. In these devices the ordering and dynamics of
ions in the interfacial layers are essential to design novel devices
for energy storage and flexible electronics.

The chemical nature of the basal planes and edges of 2D materials
determines the interactions between layers and also those with liquid
media. In graphene the extended $\pi$ electron system of $sp^2$
carbons is characterised by a high polarisability, so dispersion and
induction interactions dominate. The structure of h-BN is similar,
only finely patterned by the slightly polar \ch{B-N} bonds. TMDC such
as \ch{MoS2} are composed by three layers of atoms, covalently bonded,
in which top and bottom layers of chalcogen atoms atoms, e.g. \ch{S},
sandwich a central layer of metal atoms, e.g.\ch{Mo}. Thus there are
polar bonds in TMDC materials. In fluorographene all \ch{C} atoms are
$sp^3$ with polar \ch{C-F} bonds and the \ch{F} atoms forming a
``hard'' shell that leads to weak interactions, as in
perfluorocarbons. For different 2D materials the interactions with
solvents will be dominated by distinct terms and one objective of the
present study is to improve our understanding of how molecular and
ionic liquids organise in the interfacial layers with the materials
and how they participate in the exfoliation process.

We have been studying the non-covalent interactions of molecular and
ionic liquids with nanomaterials, trying to understant from a physical
chemistry standpoint what are the key features that determine the best
solvents for exfoliation. One of the first ideas was that the cohesive
energy densities between the solvent and the material should match
\cite{Bergin:2009dz,Coleman:2013cq}, so that liquids with a surface
tension close to the surface energy of the material should be the best
solvents. This is verified to an extent, although many solvents with
the right value of surface tension, or other solubility parameters,
prove not to be as good solvents as anticipated. The situation is
clearly more complex and our present understanding needs improvement.

Within this context we have studied exfoliation of phosphorene, which
is composed of single layers of \ch{P} atoms in a puckered structure,
with each atom bonded to three others. The \ch{P} atoms are somewhat
under-coordinated and there is a degree of covalence between layers
\cite{Sresht:2015he}, so phosphorene is not strictly speaking a van
der Waals material. We learned that solvents with flat molecules
intercalate favourably between the layers during exfoliation. This
descriptor related to molecular shape is not captured easily by
quantities such as surface tension or solubility parameters. Next we
investigated \ch{MoS2} \cite{Sresht:2017bb} and learned that the
polarity of \ch{Mo-S} bonds has a negligible effect on the contact
angle of water \cite{GovindRajan:2016bz}. These examples show that
subtle and sometimes unexpected factors play important roles.

\begin{figure}[htb]
  \centering
  \includegraphics[width=8.0cm]{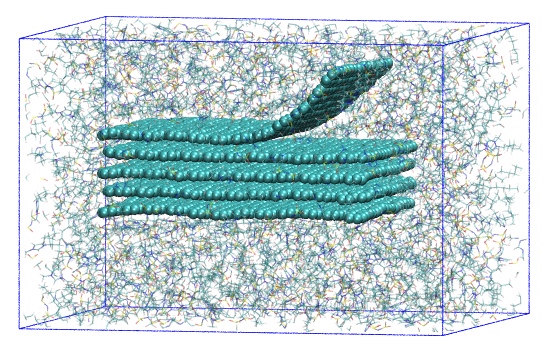}
  \includegraphics[width=8.0cm]{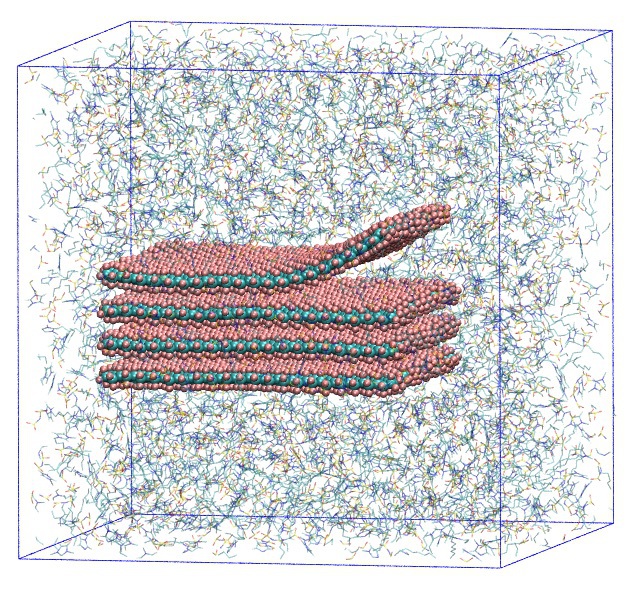}
  \caption{Simulation boxes containing a stack of 2D nanomaterial
    (graphene and fluorographene) surrounded by solvent. A total of
    about 40\,000 atoms for graphene and 60\,000 for fluorographene
    are simulated for tens of nanoseconds.}
  \label{fig:boxes}
\end{figure}

We have also recently studied the solvation of \ch{C60} fullerene and
fluorinated \ch{C60F48} in ionic liquids \cite{SzalaBilnik:2016ip},
aiming to link the chemical structure of the ions with the ability of
the ionic liquids to disaggregate and stabilise suspensions of the
nanomaterials. The present work is focused on the interactions of
molecular and ionic solvents with graphene and fluorographene, aiming
to better undertand what are the molecular features, or descriptors,
that contribute to a more efficient exfoliation of the 2D
nanomaterials. The method used is atomistic molecular simulation with
detailed interaction potentials, validated or developed specifically
for the materials studied. Classical molecular dynamics is the
adequate scale of description for the problem we wish to study,
because we need relatively large systems to represent both edges and
the basal planes of the 2D nanomaterials, in stacks and peeled
monolayers, and we also need a sufficient volume of solvent so that we
can represent both the interfacial layers and the bulk liquids. At the
same time, we wish to retain details of the interactions and so we
avoided coarse graining of the interaction models.  Typical snapshots
of the simulated systems are shown in Fig.~\ref{fig:boxes}.

Other authors have also looked at ionic liquids as dispersion and
exfoliation media for graphene or fullerene using simulation
\cite{Kamath:2012iy,Garcia:2014kv,Garcia:2015ig,Chaban:2014do,Chaban:2017bd},
but the interactions or exfoliation of fluorographene have not been
extensively studied by computational methods.

\begin{figure}[htb]
  \centering
  \includegraphics[width=7cm]{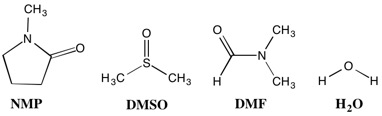}
  \caption{Chemical structures of the molecular solvents.}
  \label{fig:solvents}
\end{figure}

The molecular solvents studied here were N-methylpyrrolidone (NMP),
dimethylsulfoxide (DMSO), dimethylformamide (DMF) and water
(Fig.~\ref{fig:solvents} for molecular structures). They were chosen
because of their different interactions and functional groups, with
water being small, polar and highly associating through hydrogen
bonding, DMSO also polar, and DMF and NMP sharing the amide function
but with different molecular sizes. Amides, in particular NMP, are
among the most effective solvents to unbundle carbon nanotubes and
exfoliate graphene \cite{Coleman:2013cq}.

\begin{figure}[htb]
  \centering
  \includegraphics[width=8.5cm]{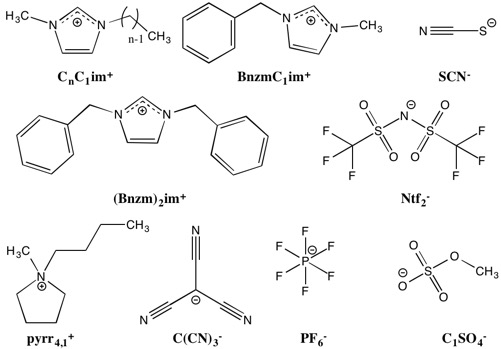}
  \caption{Chemical structures of the cations and anions of the ionic
    liquids studied here.}
  \label{fig:ions}
\end{figure}

The ionic liquids selected are based on imidazolium cations haveing
different functional groups, associated with different anions. We
explored the effect of: i) the length of the alkyl side chain on the
imidazolium cations, which determines nonpolar character, by studying
alkylmethylimidazolium cations \ch{C2C1im+}, \ch{C4C1im+} and
\ch{C10C1im+}; ii) the head group of the cation, imidazolium or
pyrrolidinium in \ch{C4C1pyr+}, the former being planar and aromatic;
and iii) the effect of aromatic benzyl groups on the imidazolium
cations, which may enhance the interaction with the extended $\pi$
system of graphene, so we studied \ch{bnzmC1im+} and
\ch{bnzm2im+}. The anions varied in size, shape and flexibility,
with some being fluorinated and others not. Anions include
hexafluorophosphate \ch{PF6-}, methysulfate \ch{C1SO4-},
bis(trifluoromethanesulfonyl)amide \ch{Ntf2-}, thiocyanate \ch{SCN-},
and tricyanomethanide \ch{C(CN)3-}. The molecular structures of the
ions studied are shown in Fig.~\ref{fig:ions}.

\section{Methods}

The structure and interactions of chemical compounds and materials
were represented by atomistic force fields, compatible with the
OPLS-AA functional form,\cite{Jorgensen:1996bs} e.g. with covalent
bonds and valence angles described by harmonic terms, torsion energy
profiles by cosine series, and non-bonded interactions by
Lennard-Jones sites and atomic partial charges. Parameters for
molecular solvents N-methylpyrrolidone (NMP), dimethylsulfoxide
(DMSO), dimethylformamide (DMF) were taken from OPLS-AA; water was
represented by the SPC/E model \cite{Berendsen:2002gc}. Ionic liquids
were modelled using the CL\&P force field
\cite{CanongiaLopes:2004he,CanongiaLopes:2012ky} with ionic charges
scaled down to $\pm0.8e$, which lead to an improved rendering of
dynamic and solvation properties \cite{Bhargava:2007cm}.  Graphene and
fluorographene sheets (and the non-covalent forces between sheets)
were described using OPLS-AA for aromatic \cite{Jorgensen:1990cn} and
perfluorinated molecules \cite{Watkins:2001ji}, respectively. The
partial charge scheme of OPLS-AA for aromatic molecules was validated
in a previous publication to be a good representation of charge
distribution in graphene planes and carbon nanotubes
\cite{Franca:-67tSnqZ}. Geometric combining rules were used for unlike
interactions between graphene or perfluorographene sheets, and also
between the molecules or ions of solvent. Between the fluorographene
and and the ionic liquids, specific interaction parameters were used
\cite{SzalaBilnik:2016ip}.

Initial configurations consisted of a stack of five graphene sheets,
or of four fluorographene sheets, placed in periodic parallelepiped
boxes using Packmol \cite{Martinez:2009di} and with the force field
generated by the fftool utility \cite{fftool:kd}. The dimensions of
the stacks are approximately \SI{5.0}{nm} by \SI{4.0}{nm} with a
thickness of \SI{1.3}{nm}. They are surrounded by sufficient solvent
in all directions (\SI{2.0}{nm} at least, after equilibration) so that
periodic images of the stacks do not affect each other, leading to
systems made of about 25\,000 atoms. Molecular dynamics (MD)
trajectories and calculations were carried out using the LAMMPS code
\cite{Plimpton:1995fc}. The timestep was \SI{1}{fs}, the cutoff for
Lennard-Jones interactions \SI{10}{\angstrom}, Coulomb interactions
handled using the PPPM method, and H-terminated bonds were constrained
using the SHAKE algorithm. Initial configurations were equilibrated at
constant temperatures of \SI{350}{K} for molecular solvents and
\SI{423}{K} for ionic liquids, and \SI{1}{bar} (regulated by
Nosé-Hoover thermostat and barostat) for \SIlist{1;5}{ns},
respectively. We chose temperatures above ambient to benefit from
higher fluidity enabling better sampling and shorter MD runs,
especially for the ionic liquids which are relatively viscous. All the
molecular solvents considered have their normal boiling points above
the chosen temperature.

\begin{figure}[htb]
  \centering
  \includegraphics[width=8.5cm]{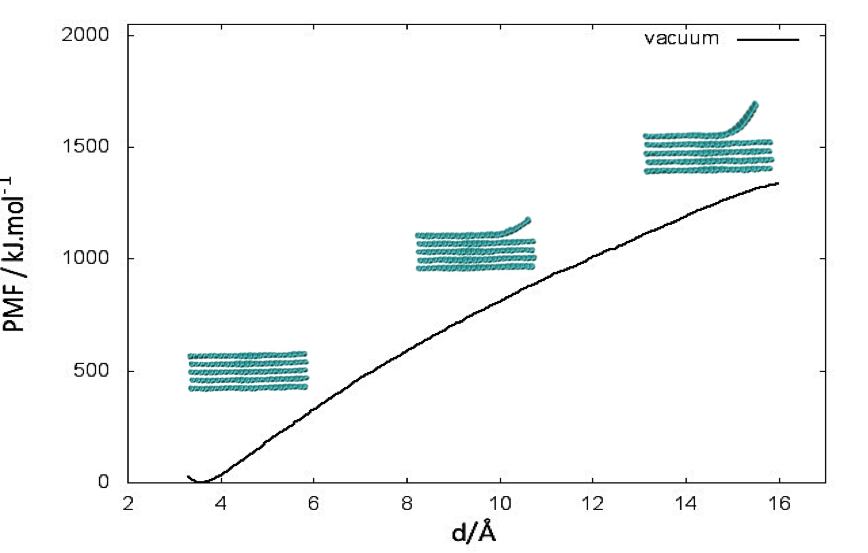}
  \caption{Calculation of the potential of mean force associated with
    peeling one layer of 2D material from a stack.}
  \label{fig:pmfcalc}
\end{figure}

The reversible work required to peel one layer of nanomaterial from
the stack was calculated via the potential of mean foce (PMF). A
perpendicular biasing potential of \SIrange{80}{180}{kJ.mol^{-1}} was
applied to the top layer of the nanomaterial, evenly distributed over
the \ch{C} atoms of its shorter edge. The \ch{C} atoms of the edge row of the
layer below the top one were tethered by a harmonic potential, and the
opposite edge of the top layer was also tethered to avoid
sliding. Except for the applied bais and tethers, the rest of the
systems evolved freely according to the flexibility of the
materials. The coordinate considered in the PMF calculations is the
distance $d$ between the centers of mass of the edge row of the top
layer and that of the edge of the layer below it. The PMF was calculated
using umbrella sampling and the weighed histogram analysis method
(WHAM). The coordinate $d$ was sampled between
\SIlist{3.0;16.0}{\angstrom} for graphene and
\SIlist{5.0;16.0}{\angstrom} for fluorographene, in steps of
\SI{0.5}{\angstrom}, and at each step the system was equilibrated for
\SI{80}{ps} followed by an acquisition period of \SI{120}{ps}. These
settings lead to a good sampling and overlap of the coordinate
histograms around each $d$ point. An illustration of how the peeling
PMF calculation proceeds is shown in Fig.~\ref{fig:pmfcalc}.

\begin{figure}[htb]
  \centering
  \includegraphics[width=8.5cm]{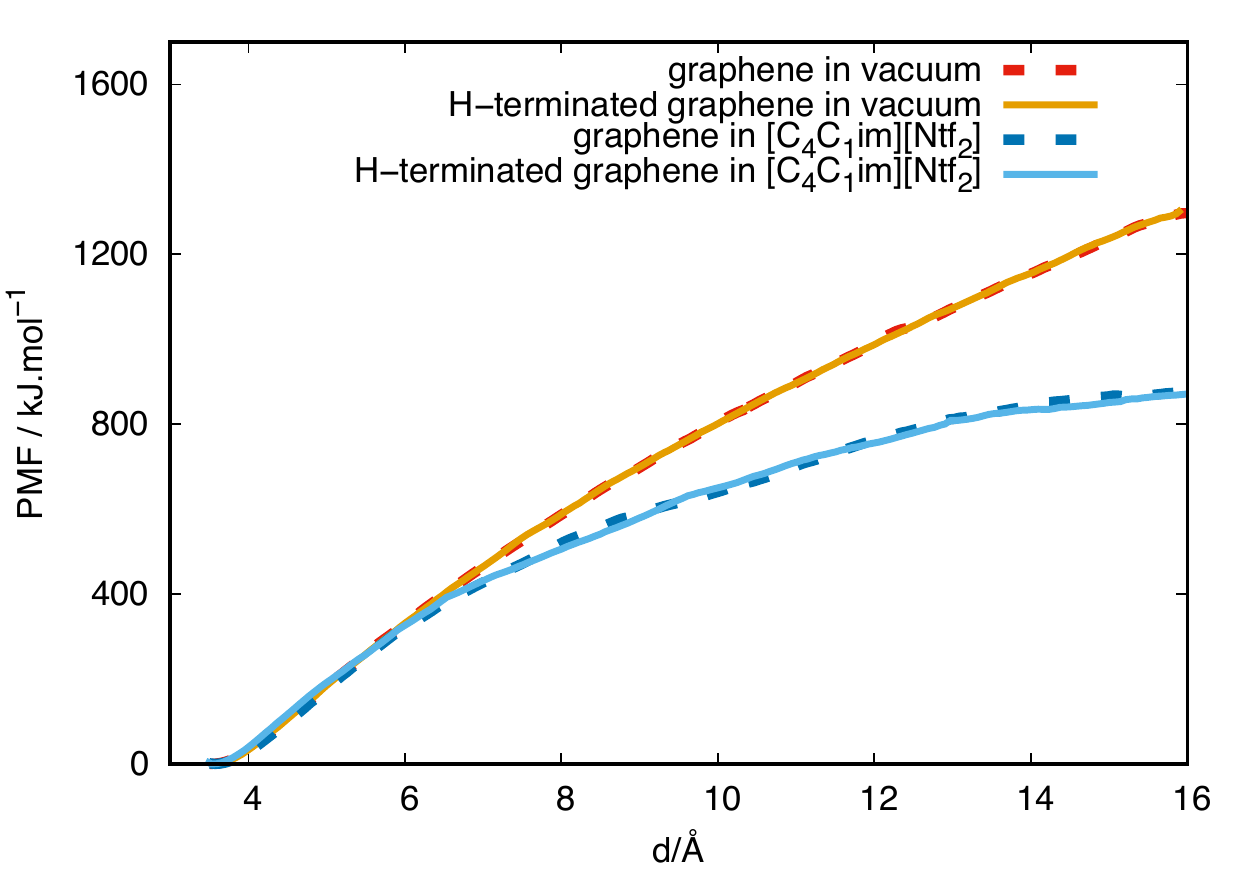}
  \caption{Effect of edge hydrogens on the PMF profile, in vacuum and
    in \ch{[C4C1im][Ntf2]} as an example.}
  \label{fig:pmfedge}
\end{figure}

The chemical state of the edges of graphene sheets may affect the
results, therefore we checked the influence of adding terminal
hydrogen atoms to all edges (zig-zag and armchair) of the graphene
sheets in our simulations. In fig.~\ref{fig:pmfedge} we compare
graphene with and without H-saturated edges, in terms of the peeling
PMF profile, both in vacuum and in one of the ionic liquids
studied. No significant difference is seen, therefore we conclude
that the present results, obtained with a simple representation
without explicit hydrogens, will be valid for H-terminated graphene
flakes as well.

The ordering of ions of the different ionic liquids near the surface
of the 2D materials was described by axial distribution functions,
$g(z) = \rho(z) / \left<\rho\right>$, where $\rho$ is the number
density of a specific atom type. These calculations were performed in
periodic boxes containing a stack of five layers of graphene or
fluorographene, periodic in the $xy$ plane. The simulated systems
contained about 20\,000 atoms and were equilibrated for \SI{10}{ns} at
\SI{423}{K} and \SI{1}{bar}. The axial density profiles were obtained
by averaging over 5000 configurations stored during \SI{5}{ns} runs.

\section{Results}

One of the main quantities reported and analysed in this work is the
PMF (free energy) associated with peeling the top layer of 2D material
from a stack. The same route is followed in vacuum and in the
solvents, so that the effect of the liquids on exfoliation can be
compared, between solvents and also with respect to vacuum. The
vaccuum calculations provide a measure of the van der Waals forces
between layers and account for the bending rigidity of the
material. We proceed with the peeling up to a certain separation,
beyond which the system enters a steady state beyond which no new
information would be obtained. Similar simulations were reported by us
in previous studies of exfoliation of other 2D materials, namely
phosphorene and \ch{MoS2}, in which the PMF method used here was set
up and validated \cite{Sresht:2015he,Sresht:2017bb}. Besides the PMF
calculations, we also analysed the ordering of the ionc liquids in the
interfacial layers with the 2D nanomaterials.

\subsection{Potential of mean force}

\begin{figure}[htb]
  \centering
  \includegraphics[width=8.5cm]{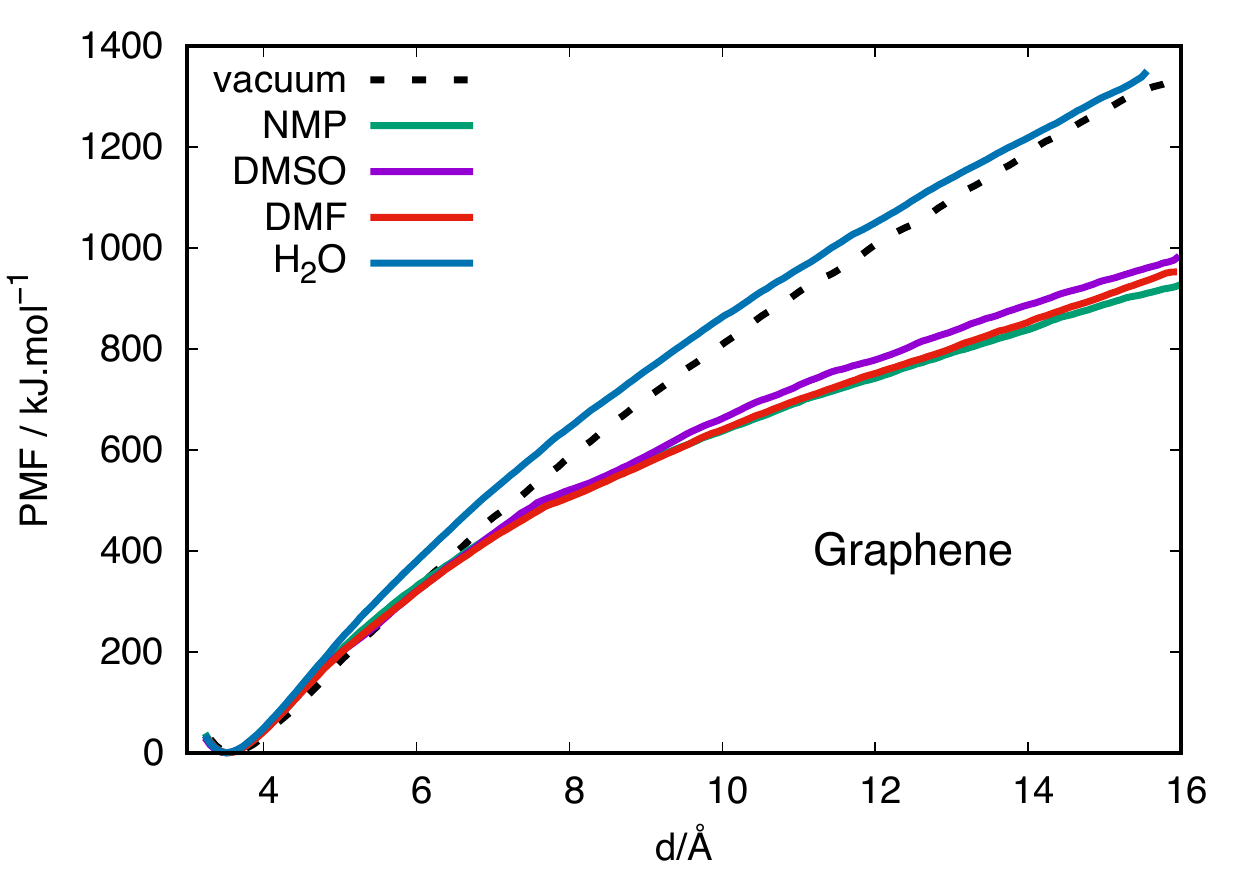}
  \includegraphics[width=8.5cm]{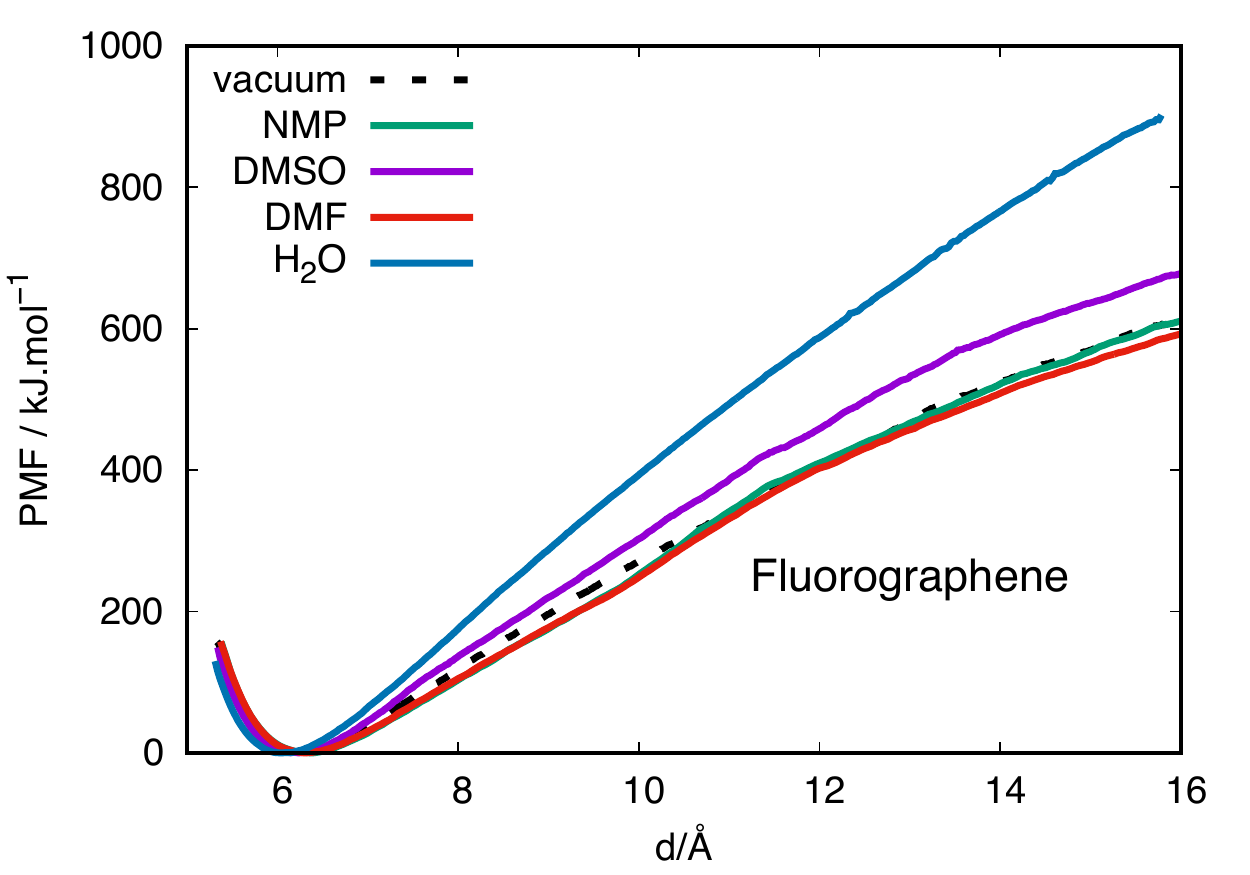}
  \caption{PMF of peeling graphene and fluorographene in molecular
    solvents at \SI{350}{K}.}
  \label{fig:pmfsolv}
\end{figure}

The PMF of peeling graphene and fluorographene in the molecular
solvents is plotted in Fig.~\ref{fig:pmfsolv}.  The reversible work
required to exfoliate graphene in vacuum is larger than the equivalent
quantity for fluorographene: at a separation of \SI{10}{\angstrom}
(counting from the equilibrium inter-layer distance) the PMF is
approximately \SI{1150}{kJ.mol^{-1}} for graphene and
\SI{600}{kJ.mol^{-1}} for fluorographene. Thus the ``fluorous''
interactions between fluorographene sheets are less attractive than
between the $\pi$ systems of graphene.

In liquids PMF curves below the vacuum curve indicate a favourable
solvent, whereas curves above the one in vacuum mean that it is more
difficult to peel away the top layer in that medium. So, lower curves
mean better solvents for exfoliation. For graphene the three organic
solvents lead to low PMF values, whereas water follows quite closely
the vacuum curve. The PMF values in the three organic solvents are
close, but distinguishable, in the order: NMP below (better solvent
than) DMF below DMSO below water. These simulation results agree with
the experimental order \cite{Coleman:2013cq}. Our previous study
\cite{SzalaBilnik:2016ip} revealed easier separation of \ch{C60} in
organic solvents like DMSO and DMF than in water, in good agreement
with the findings for graphene.

For fluorographene in the molecular solvents, PMF values stay close to
or above those in vacuum, thus the solvents studied here are not
predicted to be good for exfoliating the fluorinated 2D material.
Water is clearly the worse, followed by DMSO, with NMP and DMF values
cose to those obtained in vacuum. In relative terms, the order between
solvents is similar to that obtained for graphene.  For smaller
fluorinated objects like \ch{C60F48} \cite{SzalaBilnik:2016ip}, also
neither water nor the organic liquids tested seemed to be good
solvents.

The reason why NMP is a better exfoliation medium is is part due to
its physical solvent properties, namely that the surface tension
matches the surface energy of the material \cite{Bergin:2009dz}. This
is the argument that ``like dissolves like'': the solute-solvent
interacton energy should be commensurate with the cohesive energy of
the solvent, otherwise one of them will have too strong a tendency to
aggregate not favouring the formation of a dispersion. However, this is
not the sole descriptor defining a good solvent: the planarity of the
amides contributes to an easier intercalation between layers of the 2D
material during the exfoliation process
\cite{Sresht:2015he,Sresht:2017bb}.

\begin{figure}[htb]
  \centering
  \includegraphics[width=8.2cm]{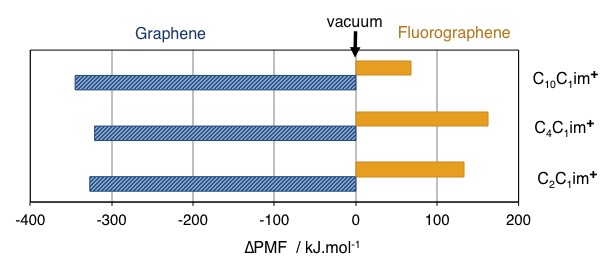}
  \includegraphics[width=8.5cm]{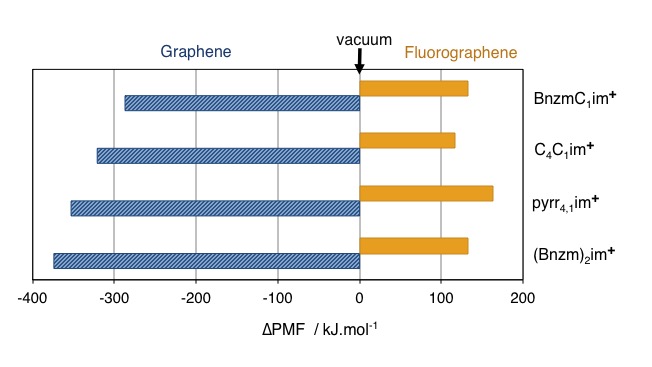}
  \caption{Differences in PMF value of peeling graphene and fluorographene
    in ionic liquids at \SI{423}{K} to a separation of
    \SI{10}{\angstrom}, with respect to peeling in vacuum. Effect of
    alkyl chain or aromatic character of the cation (all ionic liquids
    have the same \ch{Ntf2-} anion).}
  \label{fig:pmfcat}
\end{figure}

\begin{figure}[htb]
  \centering
  \includegraphics[width=8.5cm]{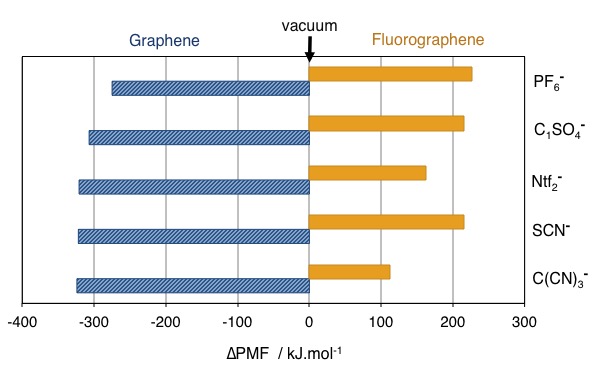}
  \caption{Differences in PMF value of peeling graphene and
    fluorographene in ionic liquids at \SI{423}{K} to a separation of
    \SI{10}{\angstrom}, with respect to vacuum. Effect of anion (all
    ionic liquids have the same \ch{C4C1im+} cation).}
  \label{fig:pmfani}
\end{figure}

PMF curves obtained in the ionc liquids are shown in
Figs.~\ref{fig:pmfcat} to \ref{fig:pmfani}. Similarly to what was
obtained in the molecular solvents, PMFs are below the vacuum curve
(favourable) for exfoliation of graphene, and above (unfavourable) for
flurographene. Different cations lead to quite comparable PMF values,
which is an interesting result. It is seen that a longer alkyl chain
is beneficial towards exfoliation, both of graphene and fluorograhene.
The pyrrolidinium head group is an improvement compared to imidazolium
(another interesting result given that imidazolium is aromatic and
thus expected to have stronger interactions with graphene). Finally,
the presence of one benzyl group is not beneficial towards graphene
exfoliation, but two benzyl groups actually lead to the lowest PMF we
calcualted. This affinity of the dibenzyl imidazolium cation towards
graphene was recently reported experimentally \cite{Bari:2014ci}.  The
best ILs for fluorographene exfoliation among those studied are based
on \ch{C10C1im+}. Also for dissolution of fullerene and fluorinated
fullerene \cite{SzalaBilnik:2016ip}, a longer alkyl side chain on the
cation was more beneficial than an aromatic group.

Changing the anion leads again to relatively small differences in
PMF of peeling graphene, and to more significant differences for
fluorographene. A ``fluorous'' effect was not observed, with
\ch{C(CN3)-} appearing to be a better anion than either \ch{PF6-} or
\ch{Ntf2-} for exfoliation of fluorographene (Fig.~\ref{fig:pmfani}).

\subsection{Interfacial layering of ionic liquids}

The ordering of ions in the interfacial layers with the 2D
nanomaterials was assessed through the axial distribution functions of
specific atoms of cations and anions,
$g(z) = \rho(z) / \left<\rho\right>$, as a function of the distance
measured parallel to the surface of the top layer. We present in what
follows some $g(z)$ of selected atoms of the ions that allow us to
extract what we consider to be significant structural features,
avoiding clutter due to the large amount of information available from
the MD trajectories.

\begin{figure}[htb]
  \centering
  \includegraphics[height=1.5cm]{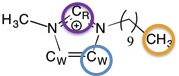}\ \ \ \ 
  \includegraphics[height=1.5cm]{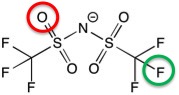}\\
  \includegraphics[width=8.5cm]{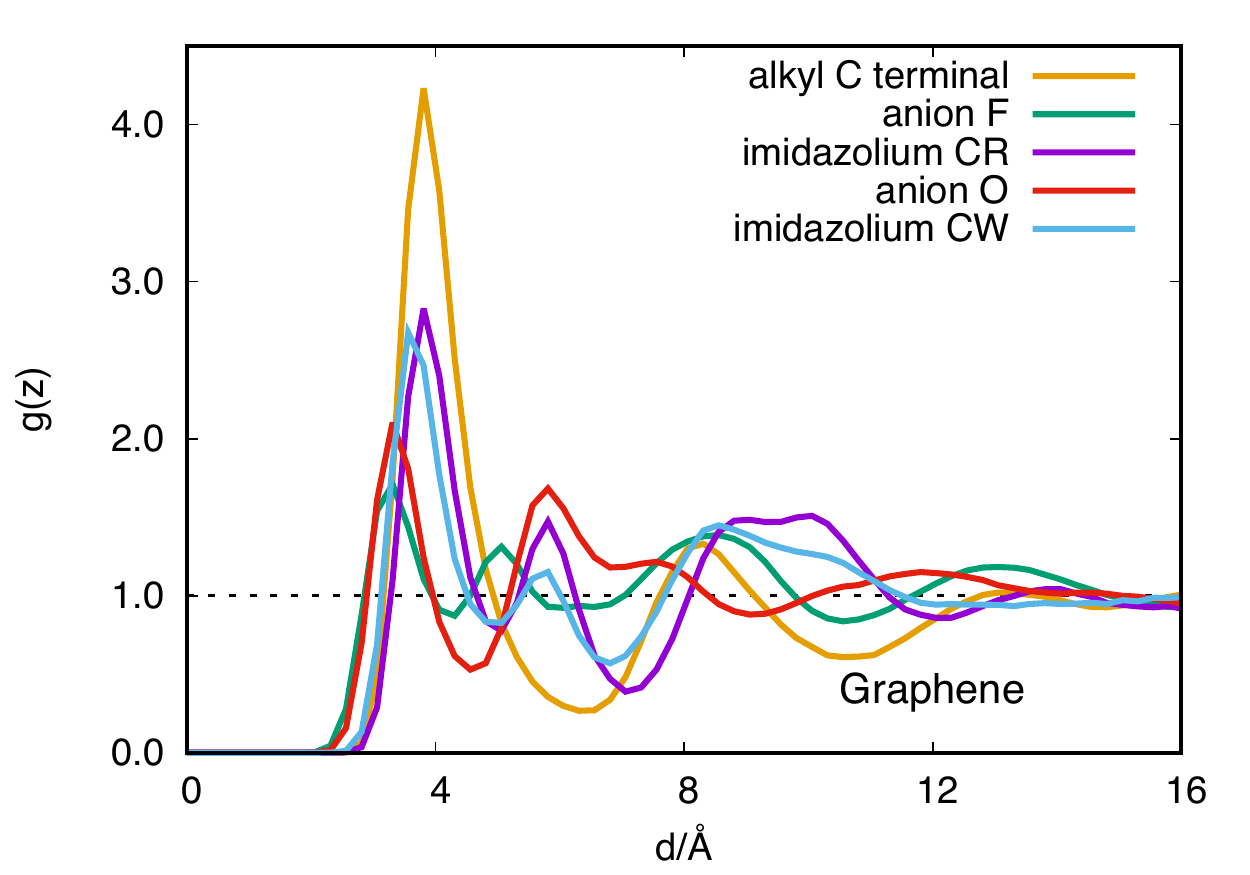}\\
  \includegraphics[width=8.5cm]{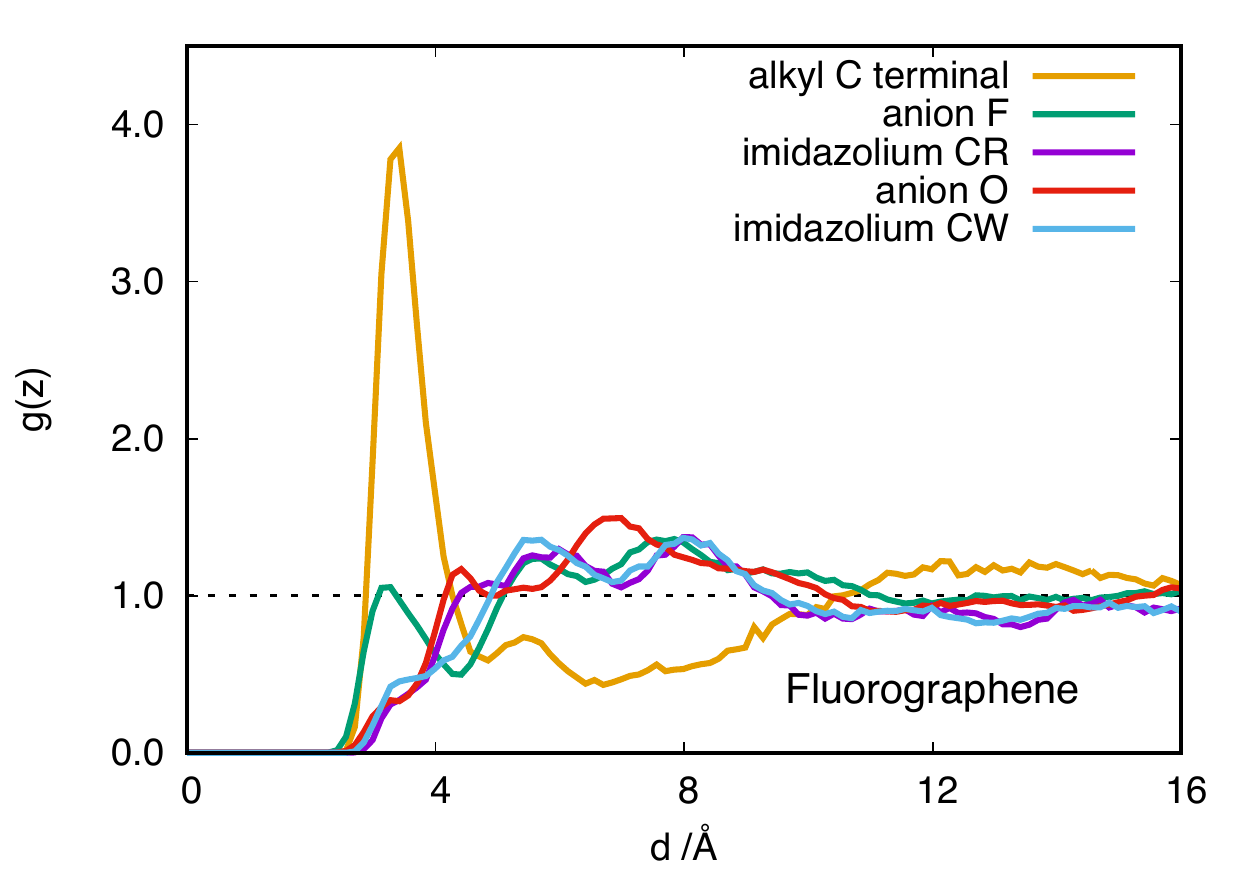}
  \caption{Distribution functions of selected atoms from
    \ch{[C10C1im][Ntf2]} as functions of distance $z$ from the surface
    of graphene and fluorogaphene.}
  \label{fig:gz-c10}
\end{figure}

The first comparison, in Fig.~\ref{fig:gz-c10}, concerns the location
of the alkyl chain of \ch{[C10C1im][Ntf2]} near the surfaces of the
materials. It is seen that the terminal \ch{C} atom of the decyl chain is
found adjacent to the surface of both materials, but near the graphene
surface are found also atoms of the imidazolium ring and of the anion,
whereas the cation headgroup and the anion show no preference for the
first interfacial layer with fluorographene. In the interfacial layer with
graphene, the \ch{C} atoms of the imidazolium ring show intense first peaks
at the same distance, indicating that the imidazolium rings are
oriented parallel to the surface. In the interfacial alyers with
fluorographene, the atoms of the cation head-group and of the anion
appear in second layer, after the alkyl chains and the \ch{F} atoms of the
anion. Fluorographene appears thus as a more hydrophobic, interacting
favourably with the alkyl side chain and the \ch{CF3} groups of the
anion, but not with the charged moieties. This apparent affinity for
fluorine atoms seen in the structural data do not translate into a
``fluorous'' effect in PMF, as discussed previously.

\begin{figure}[htb]
  \centering
  \includegraphics[height=1.5cm]{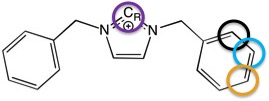}\ \ \ \ 
  \includegraphics[height=1.5cm]{ntf2}\\
  \includegraphics[width=8.5cm]{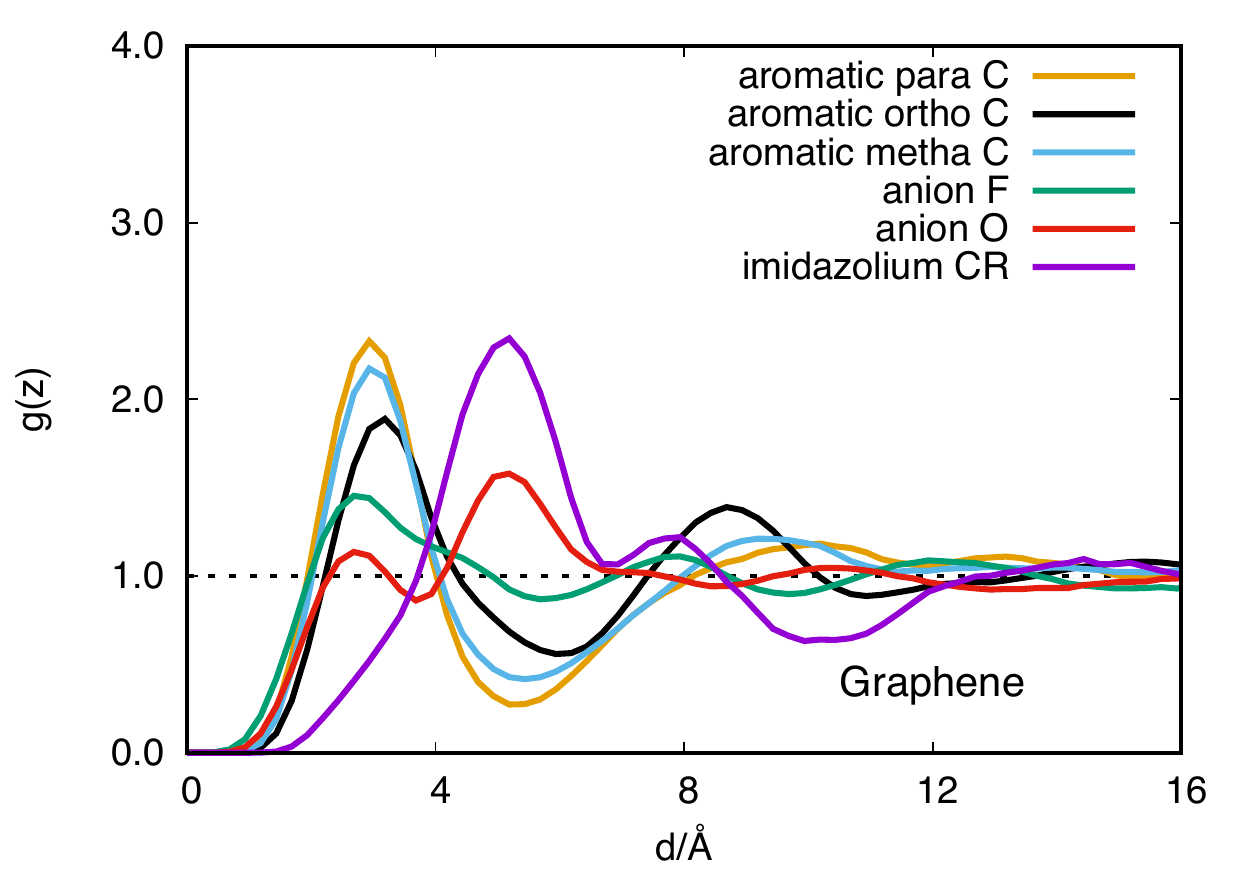}\\
  \includegraphics[width=8.5cm]{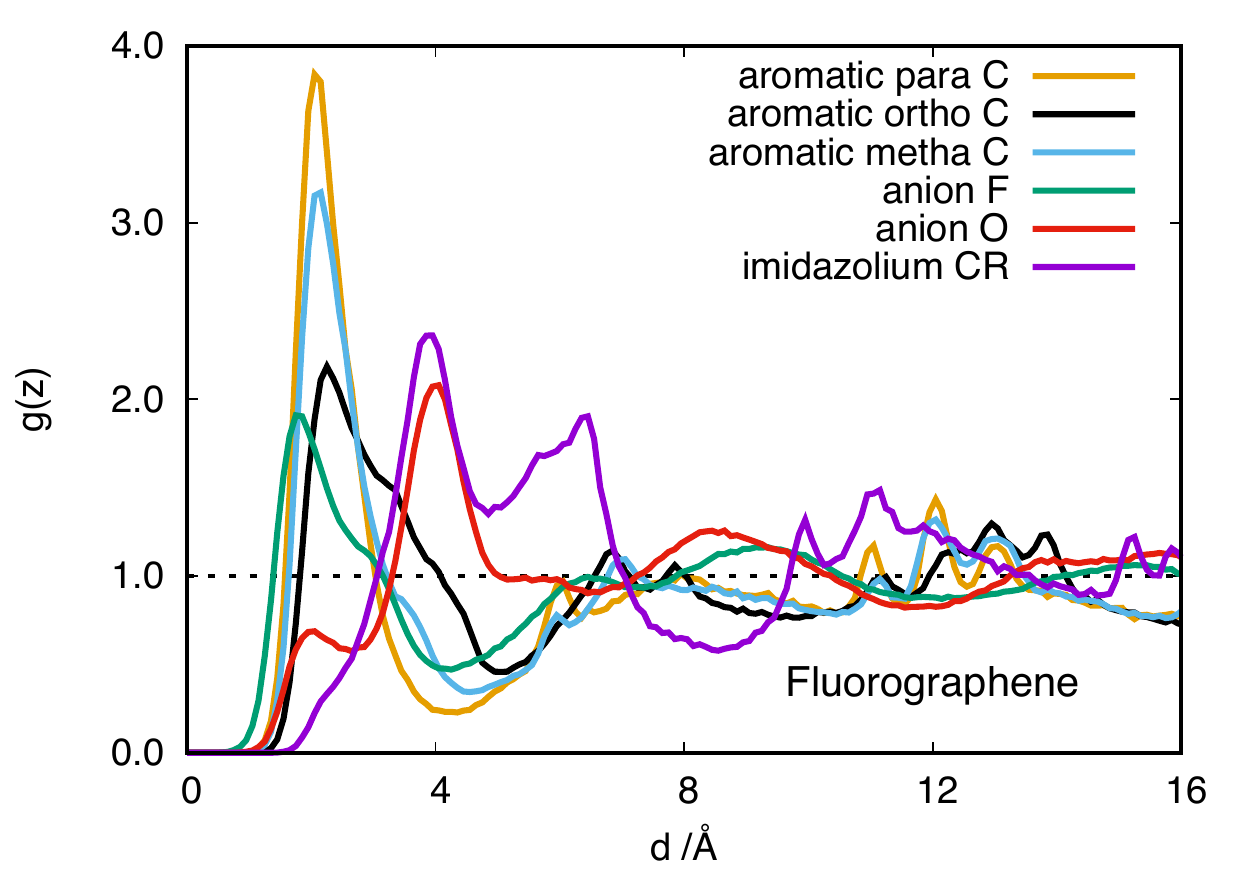}
  \caption{Distribution functions of selected atoms from
    \ch{[bnzm2im][Ntf2]} as functions of distance $z$ from the surface
    of graphene, and of \ch{[bnzmC1im][Ntf2]} with respect to fluorogaphene.}
  \label{fig:gz-benz}
\end{figure}

Next we analyse the structural features in the interfacial layers due
to the presence of benzyl groups in the imidazolium cations,
Fig.~\ref{fig:gz-benz}. We see that the first peak of different \ch{C}
atoms from the benzyl groups (ortho, meta, para) are found at the same
distance, indicating that the aromatic rings are preferentially
oriented parallel to the surface of the material. The first peaks of
the benzyl atoms are more intense and narrow near fluorographene than
near graphene. This may be due to the difference in cations, because
the imidazolium IL studied near graphene (in terms of $g(z)$) is
dibenzyl substituted, whereas the one studied with fluorographene has
only one benzyl substituent. As with the alkyl chain discussed
previously, in the interfacial layers with graphene, atoms from the
imidazolium head-group are found with higher intensity, whereas near
fluorographene only the \ch{F} atoms from the anion show significant
presence (besides the benzyl groups). Again, fluorographene leads to
more structured interfacial layers, the surface having a stronger
hydrophobic character. Contrary to what was observed with the alkyl
side chain, in the dibenzylimidazolium IL the cation head-group is not
found near the surface of graphene. Integration of the first peaks in
the $g(z)$ allows us to know the number of groups in the first
interfacial layer. For \ch{[bznm2im][Ntf2]} near graphene, the area
under the first peak is $9.4$ for the aromatic para-\ch{C}, and for
\ch{[bznmC1im][Ntf2]} the area is $4.3$, or essentially half, meaning
that the affinities of the benzyl groups for the two materials seem
comparable.

\begin{figure}[htb]
  \centering
  \includegraphics[height=1.5cm]{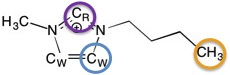}\ \ \ \ 
  \includegraphics[height=1.8cm]{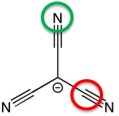}\\
  \includegraphics[width=8.5cm]{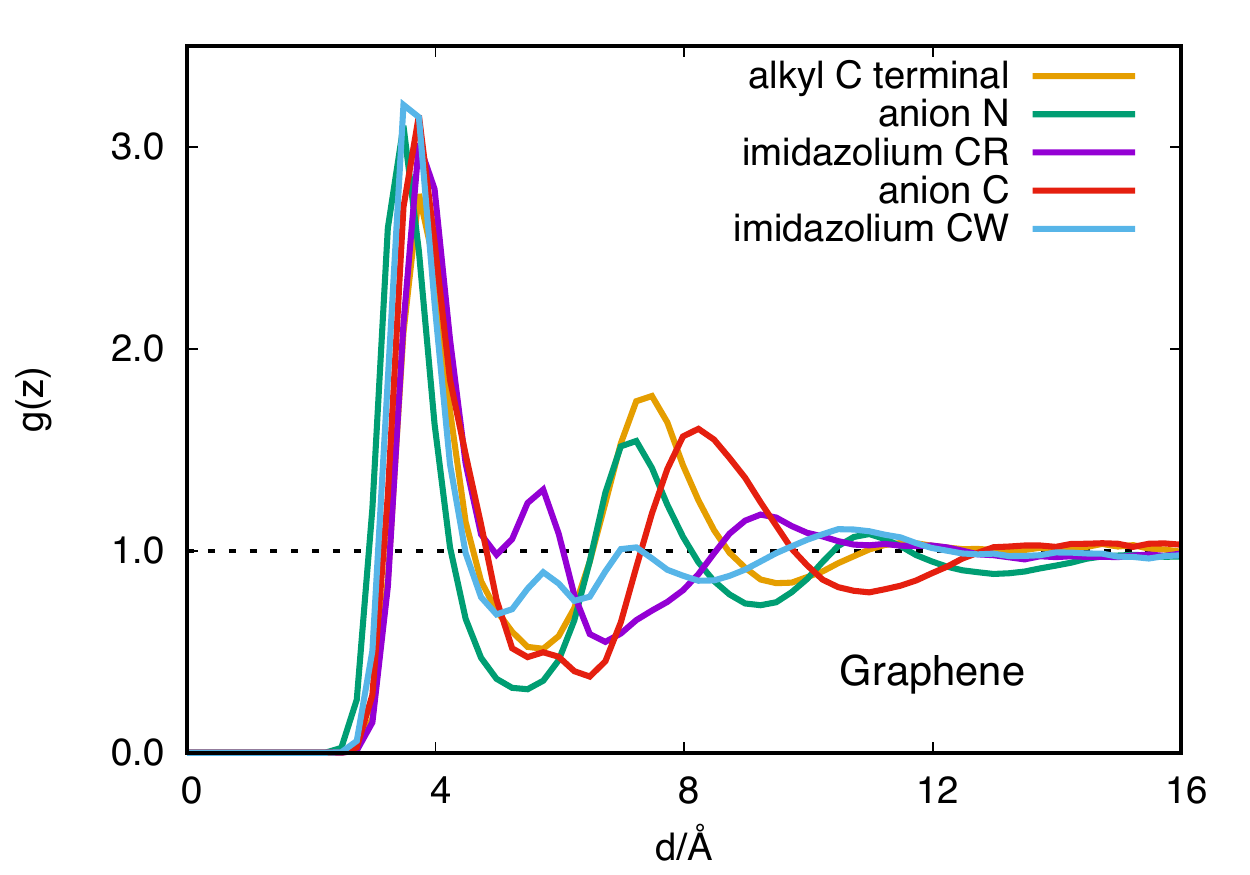}\\
  \includegraphics[width=8.5cm]{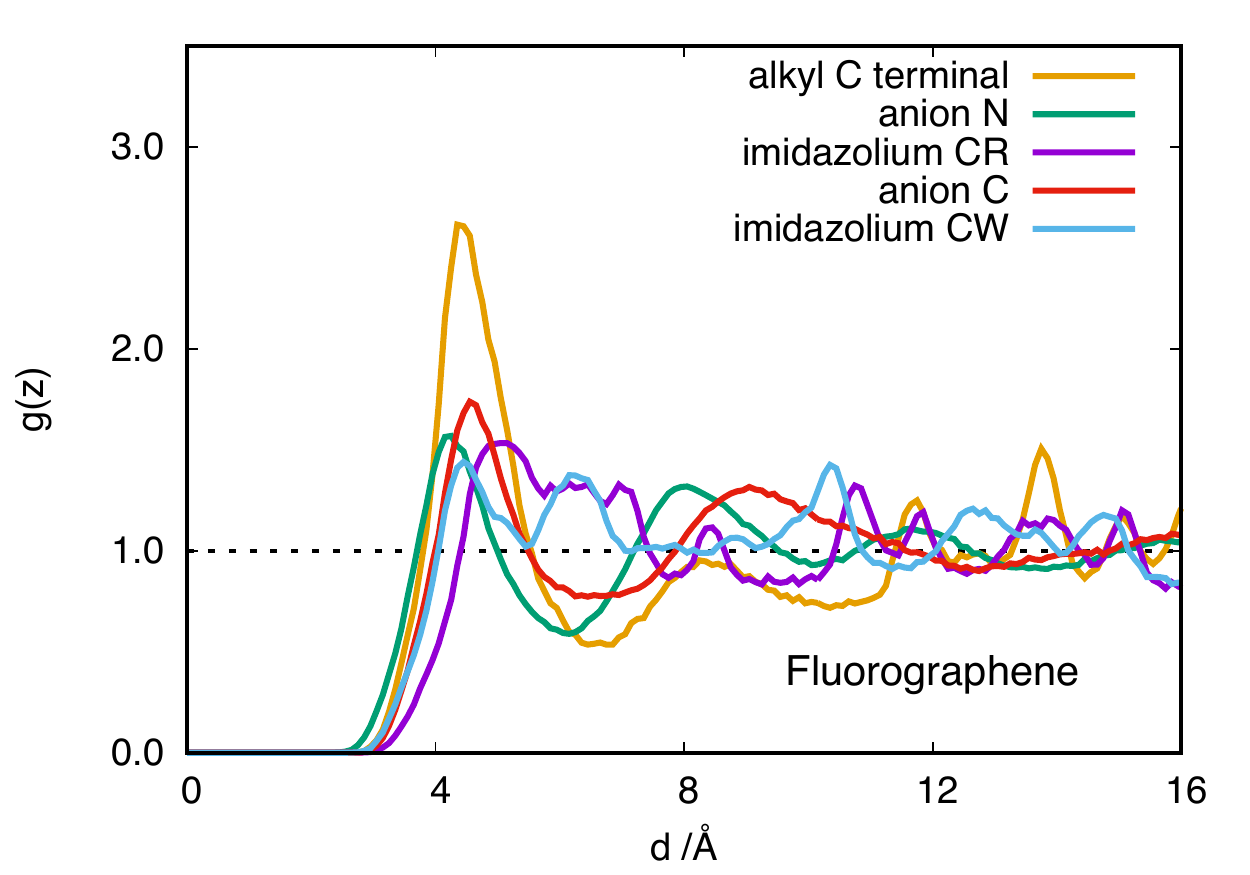}
  \caption{Distribution functions of selected atoms from
    \ch{[C4C1im][C(CN)3]} as functions of distance $z$ from the surface
    of graphene and fluorogaphene.}
  \label{fig:gz-tcm}
\end{figure}

Finally, we inspect the interfacial ordering of \ch{[C4C1im][C(CN)3]}
near the materials in Fig.~\ref{fig:gz-tcm}. It is seen that at the
interface with graphene the first peaks coincide, meaning that
imidazolium head-groups and side chains, as well as the
cyano groups from the anions, are all found in an ordered first
layer. At the fluorographene interface the situation is different,
with a predominance of terminal \ch{C} atoms from the butyl side chain
of the cation, again indicating the more hydrophobic nature of
fluorographene. Also, atoms from the cation head-group and from the
anion are found in the interfacial layer, in contrast with what was
seen in ionic liquids composed of the \ch{Ntf2-}.

\section{Conclusion}

The two types of molecular simulations we performed, to obtain the PMF
of peeling away one layer of material and to investigate the ordering
of the interfacial layers, provided several pieces of information
about molecular and ionic solvents. To start with,
attractive forces between layers of graphene are stronger than between
those of fluorographene, something that was expected.

Fluorinated graphene appears as a solvophobic material, with which
most solvents have low affinity. Graphene, on the other hand, shows
affinity for several molecular and ionic liquids, as demonstrated by
an easier peeling process in organic solvents such as NMP, and also in
ionic liquids with long side chain or aromatic functions. The order of
solvents interms of ease of exfoliation obtained here agrees with
experiment. Also, the structure at the interfacial layers is different
near the two materials: both non-polar side chains and ionic moieties
are found near graphene (more polarisable) whereas in the first layer
of ionic liquid near fluorographene mostly non-polar side chains are
found, with ionic groups displaced to second liquid layer.

In the structural data we could see \ch{CF3} groups from anions are
found near the surface of fluorographene. However, no such
``fluorous'' effect was found in the PMF results, with non-halogenated
anions such as \ch{C(CN)3-} proving to be the best for
exfoliation. Another aspect of ``like dissolves like'' that we tested
was the effect of aromatic substituent groups in the cations.
Dibenzyl imidazolium was found to be a favourable cation for graphene,
although not for fluorographene.

Overall, with the families of ionic liquids we studied here, the
effect of modifying the cation led to more important changes in PMF
than the choice of anion.

\section{Acknowledgments}

This work was supported by the Agence Nationale de la Recherche project CLINT
ANR-12-IS10-003.

%

\end{document}